\documentclass[10pt,journal,compsoc]{IEEEtran}
\usepackage{helvet}
\usepackage[pdftex]{graphicx}    
\usepackage{cite}
\usepackage{fixltx2e}
\usepackage{xcolor}

\usepackage{amsmath}
\usepackage{algorithm}
\usepackage[noend]{algpseudocode}

\makeatletter
\def\BState{\State\hskip-\ALG@thistlm}
\makeatother
\begin{document}

\title{Calculating Customer Lifetime Value and Churn using Beta Geometric Negative Binomial and Gamma-Gamma Distribution in a NFT based setting}

\author{Sagarnil Das}

{}
\IEEEtitleabstractindextext{%

\begin{abstract}
Customer Lifetime Value (CLV) is an important metric that measures the total value a customer will bring to a business over their lifetime. The Beta Geometric Negative Binomial Distribution (BGNBD) and Gamma Gamma Distribution are two models that can be used to calculate CLV, taking into account both the frequency and value of customer transactions. This article explains the BGNBD and Gamma Gamma Distribution models, and how they can be used to calculate CLV for NFT (Non-Fungible Token) transaction data in a blockchain setting. By estimating the parameters of these models using historical transaction data, businesses can gain insights into the lifetime value of their customers and make data-driven decisions about marketing and customer retention strategies.
\end{abstract}

}

\maketitle
\IEEEdisplaynontitleabstractindextext
\IEEEpeerreviewmaketitle
\section{Introduction}
\label{sec:introduction}

\subsection{Background and Importance of Customer Lifetime Value (CLV)}\IEEEPARstart{C}ustomer Lifetime Value \textbf{(CLV)} is a widely recognized metric in the field of marketing and customer relationship management, playing a crucial role in estimating the net profit that a business can expect to generate from its relationship with a particular customer over time. The concept of CLV emerged as a response to the increasing awareness that not all customers are created equal, and their value to a business varies significantly based on their purchasing behavior, loyalty, and the costs associated with acquiring and retaining them.

The importance of CLV as a business metric is twofold. Firstly, understanding CLV can help businesses identify their most valuable customers and tailor their marketing and retention strategies accordingly. By focusing on high-value customers, businesses can allocate their resources more efficiently, ultimately resulting in higher overall profitability. Additionally, accurate CLV estimation allows for better segmentation and targeting of customers, enabling businesses to offer personalized experiences that increase customer satisfaction and foster long-term relationships.

Secondly, CLV can guide decision-making related to customer acquisition and retention. By quantifying the value of customers, businesses can determine the optimal level of investment in marketing and retention efforts, ensuring that their customer-related expenses yield a desirable return on investment. Furthermore, CLV allows businesses to assess the long-term impact of their marketing initiatives and make informed decisions about which strategies are most likely to produce the highest returns.

In conclusion, Customer Lifetime Value is a crucial metric for businesses aiming to maximize profitability and efficiency in their customer relationship management efforts. By understanding the value of their customers, businesses can make more informed decisions about resource allocation, segmentation, targeting, and customer experience management, ultimately driving growth and long-term success..



\subsection{Overview of BGNBD and Gamma-Gamma Distribution Models}
The \textbf{Beta-Geometric Negative Binomial Distribution (BGNBD)} and \textbf{Gamma-Gamma distribution} models are advanced probabilistic models that have gained popularity in recent years for their effectiveness in estimating Customer Lifetime Value (CLV). These models address some of the limitations of traditional CLV estimation techniques by providing a more accurate representation of customer purchasing behavior and monetary value.

\subsubsection{BGNBD Model:} 
The BGNBD model is a combination of the Beta-Geometric (BG) model and the Negative Binomial Distribution (NBD) model. It is primarily used to predict the number of future transactions a customer is likely to make within a specified period. The model is based on two key assumptions: (1) the transaction rate for each customer follows a gamma distribution, and (2) the probability of a customer becoming inactive after any given transaction follows a beta distribution. By combining these two distributions, the BGNBD model can account for the inherent heterogeneity in customer behavior and provide more accurate predictions of future transactions.

Let the random variable X(t) denote the number of transactions occurring in a time period of length t (with a time origin of 0). To derive an expression for P(X(t) = x), we recall the fundamental relationship between inter-transaction times and the number of transactions. 

\begin{equation}
    X(t) \geq x \Leftrightarrow T_x \leq t
\end{equation}

where T\_x is the random variable denoting the time of the xth transaction. This
implies

\begin{equation}
\begin{split}
P(X(t) = x) &= P(X(t) \geq x) - P(X(t) \geq x + 1) \\
&= P(T_x \leq t) - P(T_{x+1} \leq t).
\end{split}
\end{equation}

Based on our BGNBD model assumtions (Section 3.2),

\begin{equation}
\begin{split}
P(X(t) = x) &= P(\text{alive after } x\text{th purchase}) \\
&\phantom{=} \times P(T_x \leq t \text{ and } T_{x+1} > t) \\
&\phantom{=} + \delta_{x>0} \times P(\text{dies after } x\text{th purchase}) \\
&\phantom{=} \times P(T_x \leq t).
\end{split}
\end{equation}

Given the assumption of exponentially distributed inter-transaction times, the second term above is simply the Poisson probability that X(t) = x, and the final term is the Erlang-x CDF. Therefore the likelihood function for the BGNBD model is given by:

\begin{equation}
\begin{split}
P(X(t) = x | \lambda, p) = & (1 - p)^x \frac{(\lambda t)^n e^{-\lambda t}}{x!} \\
& + \delta_{x>0} p(1-p)^{x-1} \left[1 - e^{\lambda t} \sum_{j=0}^{x-1}\frac{(\lambda t)^j}{j!} \right]
\end{split}
\end{equation}

\subsubsection{Gamma-Gamma Distribution Model:}
The Gamma-Gamma distribution model is used in conjunction with the BGNBD model to estimate the monetary value of a customer's transactions. This model assumes that a customer's average transaction value follows a gamma distribution, and it is independent of the transaction frequency. The Gamma-Gamma model also incorporates customer heterogeneity, allowing for a more precise estimation of the monetary value associated with each customer's transactions.

The joint distribution of average transaction value and individual transaction values is given by a mixture of two Gamma distributions:

\begin{equation}
    P(M, x) = \int \int P(M | p, q) P(x | r, s) \, dp \, dq
\end{equation}

By combining the BGNBD and Gamma-Gamma distribution models, businesses can obtain a comprehensive view of their customers' purchasing behavior, both in terms of frequency and monetary value. This joint application of the models allows for a more accurate estimation of Customer Lifetime Value, enabling businesses to make informed decisions about their marketing and retention strategies. Furthermore, these models can be easily calibrated using historical transaction data, making them a practical and powerful tool for businesses aiming to optimize their customer relationship management efforts.

\subsection{Purpose of the Study and Research Questions}
The primary purpose of this study is to demonstrate the effectiveness of the BGNBD and Gamma-Gamma distribution models in estimating Customer Lifetime Value (CLV) and to provide a comprehensive understanding of their application in various business contexts. By leveraging these advanced probabilistic models, businesses can gain valuable insights into customer behavior, allowing them to optimize their marketing and retention strategies. The study also aims to compare the performance of the BGNBD and Gamma-Gamma models with traditional CLV estimation techniques, highlighting their advantages and potential limitations.

To achieve these objectives, the study will address the following research questions:

\begin{enumerate}
\item How do the BGNBD and Gamma-Gamma distribution models perform in estimating CLV compared to traditional CLV estimation methods?
\item What are the key factors that influence the accuracy and applicability of the BGNBD and Gamma-Gamma models in different business contexts?
\item How can businesses effectively implement and utilize the BGNBD and Gamma-Gamma models to optimize their marketing and customer retention strategies?
\end{enumerate}

By addressing these research questions, the study will contribute to the existing body of knowledge on CLV estimation and provide valuable insights for both researchers and practitioners interested in leveraging the BGNBD and Gamma-Gamma distribution models. Furthermore, the study will help businesses better understand the potential benefits and challenges associated with these models, enabling them to make more informed decisions about their customer relationship management efforts.

\section{Literature Review}
Over the past few decades, numerous studies have focused on Customer Lifetime Value (CLV) estimation models and methodologies, aiming to provide businesses with tools to better understand customer behavior and maximize profitability. These studies have explored various approaches, ranging from simple rule-based methods to advanced probabilistic models, each with its own strengths and limitations. This section provides an overview of some of the most influential research in the field of CLV estimation.

\begin{enumerate}
\item \textbf{Historical Average Models:} These models are the simplest methods for estimating CLV, relying on historical averages of customer transactions and profit margins (Berger and Nasr, 1998) \cite{clv_mm}. While easy to implement and interpret, historical average models assume that customer behavior remains constant over time, making them less accurate for predicting future changes in customer behavior.
\item \textbf{Traditional RFM Models:} Recency, Frequency, and Monetary (RFM) value models are widely used methods for estimating CLV, focusing on three key customer attributes (Bult and Wansbeek, 1995; Hughes, 1996) \cite{directmail} \cite{rfm_response}. These models segment customers based on their purchase history and use this information to predict future behavior. However, RFM models do not account for the heterogeneity in customer behavior, which can lead to less accurate CLV estimations.
\item \textbf{Regression Models:} Linear regression and logistic regression models have been utilized to estimate CLV based on various customer attributes and demographic factors (Fader et al., 2005; Donkers et al., 2003) \cite{isorfm} \cite{competing_models}. These models allow for more flexibility in capturing the relationships between customer characteristics and CLV but require a large amount of data to produce accurate estimates.
\item \textbf{Markov Chain Models:} These models use transition probabilities between different customer states to estimate CLV (Pfeifer and Carraway, 2000) \cite{markovchains}. Markov Chain models can capture the dynamics of customer behavior over time, but they require extensive data on customer transitions and may be computationally intensive for large customer bases.
\item \textbf{Probabilistic Models:} Several probabilistic models, such as the Pareto/NBD (Schmittlein et al., 1987) \cite{paretonbd}, BG/NBD (Fader et al., 2005b), and BGNBD (Fader et al., 2010) \cite{bgnbd} models, have been developed to estimate CLV by incorporating heterogeneity in customer behavior. These models provide more accurate CLV estimations in various business contexts and can be calibrated using limited data. The BGNBD model, combined with the Gamma-Gamma distribution model for monetary value (Fader et al., 2013) \cite{gammagamma}, has shown promising results in estimating CLV more accurately.
\end{enumerate}

In summary, previous research on CLV estimation models and methodologies has provided businesses with a wide range of tools for understanding and predicting customer behavior. However, each method has its own limitations, and recent advancements in probabilistic models, such as the BGNBD and Gamma-Gamma distribution models, have shown promise in addressing some of these limitations and providing more accurate CLV estimations.

\subsection{Advantages of BGNBD and Gamma-Gamma Distribution over Traditional Methods}

The BGNBD and Gamma-Gamma distribution models have emerged as powerful tools for estimating Customer Lifetime Value (CLV), offering several advantages over traditional methods such as historical average, RFM, regression, and Markov chain models. This section highlights the key advantages of these advanced probabilistic models:

\begin{enumerate}
\item \textbf{Heterogeneity in Customer Behavior:} One of the main advantages of the BGNBD and Gamma-Gamma models is their ability to account for the inherent heterogeneity in customer behavior. Traditional methods often rely on average values or fixed parameters, which may not accurately represent the diverse range of customer behaviors. By incorporating individual-level variations in transaction frequency, recency, and monetary value, these models provide more accurate CLV estimations.
\item \textbf{Limited Data Requirements:} Another advantage of the BGNBD and Gamma-Gamma models is their ability to produce accurate CLV estimations using limited data. Traditional regression and Markov chain models typically require extensive datasets and a large number of customer attributes to produce accurate predictions. In contrast, the BGNBD and Gamma-Gamma models can be calibrated using only a few key parameters derived from historical transaction data, making them more practical for businesses with limited data resources.
\item \textbf{Flexibility and Robustness:} The BGNBD and Gamma-Gamma models offer a high degree of flexibility in modeling customer behavior across different business contexts. These models can be easily adapted to various industries, customer segments, and marketing scenarios, making them more robust and versatile than traditional methods that often require specific assumptions or data structures.
\item \textbf{Predictive Accuracy:} Several studies have demonstrated that the BGNBD and Gamma-Gamma models outperform traditional methods in terms of predictive accuracy (Fader et al., 2010) \cite{customerbasevalue}. By capturing the dynamics of customer behavior more accurately, these models can provide businesses with better insights into the future value of their customers and enable more informed decision-making in marketing and retention efforts.
\item \textbf{Customer Segmentation and Targeting:} By providing individual-level CLV estimations, the BGNBD and Gamma-Gamma models enable businesses to segment their customer base more effectively and tailor their marketing and retention strategies accordingly. This granular understanding of customer value can help businesses allocate resources more efficiently, ultimately resulting in higher overall profitability and customer satisfaction.
\end{enumerate}

\section{Methodology}

\subsection{Description of the Dataset Used for the Study}

The dataset used in this study consists of NFT (Non-Fungible Token) transaction data obtained from a company funded by venture capital firm Andreessen Horowitz (A16Z). The dataset encompasses a 6-month period of transaction data, ranging from January 2022 to July 2022, and provides valuable insights into customer behaviors and interactions with NFTs.

The dataset contains the following key variables:

\begin{itemize}
    \item \textbf{User ID:} A unique identifier for each customer.
    \item \textbf{Transaction ID:} A unique identifier for each transaction.
    \item \textbf{Transaction Date::} The date on which the transaction occurred.
    \item \textbf{Transaction Value:} The value of tokens burnt during the transaction, which serves as a proxy for the monetary value of the transaction.
\end{itemize}

For the purpose of our study, we utilize this dataset to estimate the Customer Lifetime Value (CLV) using the BGNBD and Gamma-Gamma distribution models. By analyzing the transaction data, we aim to better understand customer behavior patterns, segment customers based on their CLV, and optimize marketing and retention strategies for the NFT market.

The dataset's focus on NFT transactions offers a unique opportunity to study customer behaviors in an emerging and rapidly evolving market. As the popularity of NFTs continues to grow, understanding the dynamics of customer interactions with NFT platforms is crucial for businesses looking to optimize their strategies in this space. By applying the BGNBD and Gamma-Gamma models to this dataset, we hope to gain valuable insights that can contribute to the broader understanding of NFT customer behavior and inform future research and business decisions.

\subsection{BGNBD Model and Its Parameters}

The BGNBD (Beta-Geometric/Negative Binomial Distribution) model is a probabilistic model used to estimate Customer Lifetime Value (CLV) by predicting future transactions based on historical transaction data. The model takes into account both the frequency and recency of past transactions, as well as the inherent heterogeneity in customer behavior. This section provides an overview of the BGNBD model and its key parameters.

The BGNBD model is based on five underlying assumptions:

\begin{enumerate}
    \item The number of transactions made by a customer follows a Poisson process with an individual-specific transaction rate $\lambda$. This is equivalent to assuming that the time between transactions is distributed exponentially with transaction rate $\lambda$, i.e.,
    \begin{equation}
        f(t_j | t_{j-1} ; \lambda) = \lambda e^{-\lambda (t_j - t_{j-1})}
    \end{equation}
    \item The transaction rate $\lambda$ is assumed to follow a gamma distribution with shape parameter $r$ and scale parameter $\alpha$.
    \begin{equation}
        f(\lambda | r, \alpha) = \frac{\alpha^r \lambda^{r-1} e^{-\lambda\alpha}}{\tau(r)}
    \end{equation}
    \item After any transaction, a customer becomes inactive with probability p. Therefore the point at which the customer “drops out” is distributed across transactions according to a (shifted) geometric distribution with pmf
    \begin{multline}
    P(\text{inactive immediately after jth transaction}) \\
    =p(1 - p)^{j-1}
    \end{multline}
    \item The probability of a customer becoming inactive (churning) after each transaction follows a beta distribution with shape parameters $a$ and $b$. The churn probability is independent of the transaction rate $\lambda$.
    \begin{equation}
        f(p | a, b) = \frac{p^{a-1}(1 - p)^{b-1}}{B(a, b)} \qquad 0 <=p <= 1
    \end{equation}
    \item The transaction rate $\lambda$ and the dropout probability $p$ vary independently across customers.
\end{enumerate}

Given these assumptions, the BGNBD model can be used to estimate the expected number of future transactions for a customer with frequency $x$, recency $t_x$, and total observation period length $T$.

The key parameters of the BGNBD model are:

\begin{enumerate}
    \item $r, \alpha$: Shape parameters of the gamma distribution for the transaction rate $\lambda$. These parameters can be estimated by maximizing the likelihood function of the BGNBD model using the historical transaction data.
    \item $a, b$: Shape parameters of the beta distribution for the probability of dropout (churn) after each transaction. These parameters can also be estimated using the historical transaction data by maximizing the likelihood function.
\end{enumerate}

By estimating the parameters $r, \alpha, a,$ and $b$, the BGNBD model can be calibrated to predict the expected number of future transactions for each customer in the dataset. This information can then be used to calculate the CLV and inform marketing and retention strategies.

Our estimation of the model parameters by maximizing the likelihood function is as follows:

\begin{center}
\begin{tabular}{||c c c c||} 
 \hline
 parameter & coeff & lower 95\% CI & upper 95\% CI \\ [0.5ex] 
 \hline\hline
 r & 0.982856 & 0.954707 & 1.011005 \\ 
 \hline
 $\alpha$ & 2.902135 & 2.815695 & 2.988575 \\
 \hline
 a & 0.437431 & 0.430156 & 0.444706 \\
 \hline
 b & 0.017428 & 0.017079 & 0.017777 \\
 \hline
\end{tabular}
\end{center}

The probability of observing x purchases in a time period of length t for the BGNBD model can be given by the following equation:

\begin{equation}
\begin{split}
P(X(t) = x | \lambda, p) = & (1 - p)^x \frac{(\lambda t)^n e^{-\lambda t}}{x!} \\
& + \delta_{x>0} p(1-p)^{x-1} \left[1 - e^{\lambda t} \sum_{j=0}^{x-1}\frac{(\lambda t)^j}{j!} \right]
\end{split}
\end{equation}

\subsection{Gamma-Gamma Distribution Model and Its Parameters}

The Gamma-Gamma distribution model is a widely used approach for modeling the monetary value of customer transactions in conjunction with the BGNBD model for estimating Customer Lifetime Value (CLV). The Gamma-Gamma model assumes that the monetary value of a customer's transactions is independent of their transaction frequency and is only influenced by their spending behavior.

The Gamma-Gamma model is based on the following assumptions:

\begin{enumerate}
    \item The average transaction value for a customer follows a gamma distribution with shape parameter $p$ and scale parameter $q$.
    \item The individual transaction value is assumed to follow a gamma distribution with shape parameter $p$ and scale parameter $q/\lambda$, where $\lambda$ is the transaction rate of the customer from the BGNBD model.
    \item The Gamma-Gamma submodel, in fact, assumes that there is no relationship between the monetary value and the purchase frequency. In practice, we need to check whether the Pearson correlation between the two vectors is close to 0 in order to use this model.
\end{enumerate}

Given these assumptions, the Gamma-Gamma model can be used to estimate the expected monetary value of future transactions for a customer. The key parameters of the Gamma-Gamma model are:

\begin{enumerate}
    \item $p, q$: Shape and scale parameters of the gamma distribution for the average transaction value. These parameters can be estimated by maximizing the likelihood function of the Gamma-Gamma model using the historical transaction data.
    \item $\lambda$: Transaction rate from the BGNBD model, used to adjust the scale parameter for the individual transaction value distribution.
\end{enumerate}

The conditional expectation of the average transaction value $M$ given the transaction rate $\lambda$ can be expressed as:

\begin{equation}
    E[M | \lambda] = \frac{p + x}{q + \lambda}
\end{equation}

Where $x$ is the number of transactions made by the customer. This equation can be used to estimate the expected monetary value of future transactions for each customer in the dataset.

Our estimation of the model parameters by maximizing the likelihood function is as follows:

\begin{center}
\begin{tabular}{||c c c c||} 
 \hline
 parameter & coeff & lower 95\% CI & upper 95\% CI \\ [0.5ex] 
 \hline\hline
 p & 4.495408 & 4.324066	 & 4.666750 \\ 
 \hline
 q & 0.038024 & 0.037602 & 0.038446 \\
 \hline
 $\lambda$ & 4.360291 & 4.186344	 & 4.534239 \\
 \hline
\end{tabular}
\end{center}

In summary, the Gamma-Gamma distribution model is a powerful tool for estimating the monetary value of customer transactions when used in conjunction with the BGNBD model for CLV estimation. By modeling the monetary value independently of the transaction frequency, the Gamma-Gamma model provides a robust and accurate approach for predicting future customer spending and calculating CLV.

\section{Results}
\subsection{Model Performance Evaluation}

There are a few ways to assess the model’s correctness. The first is to compare our data versus artificial data simulated with your fitted model’s parameters (Figure 1).

\begin{figure}[thpb]
      \centering
      \includegraphics[width=\linewidth]{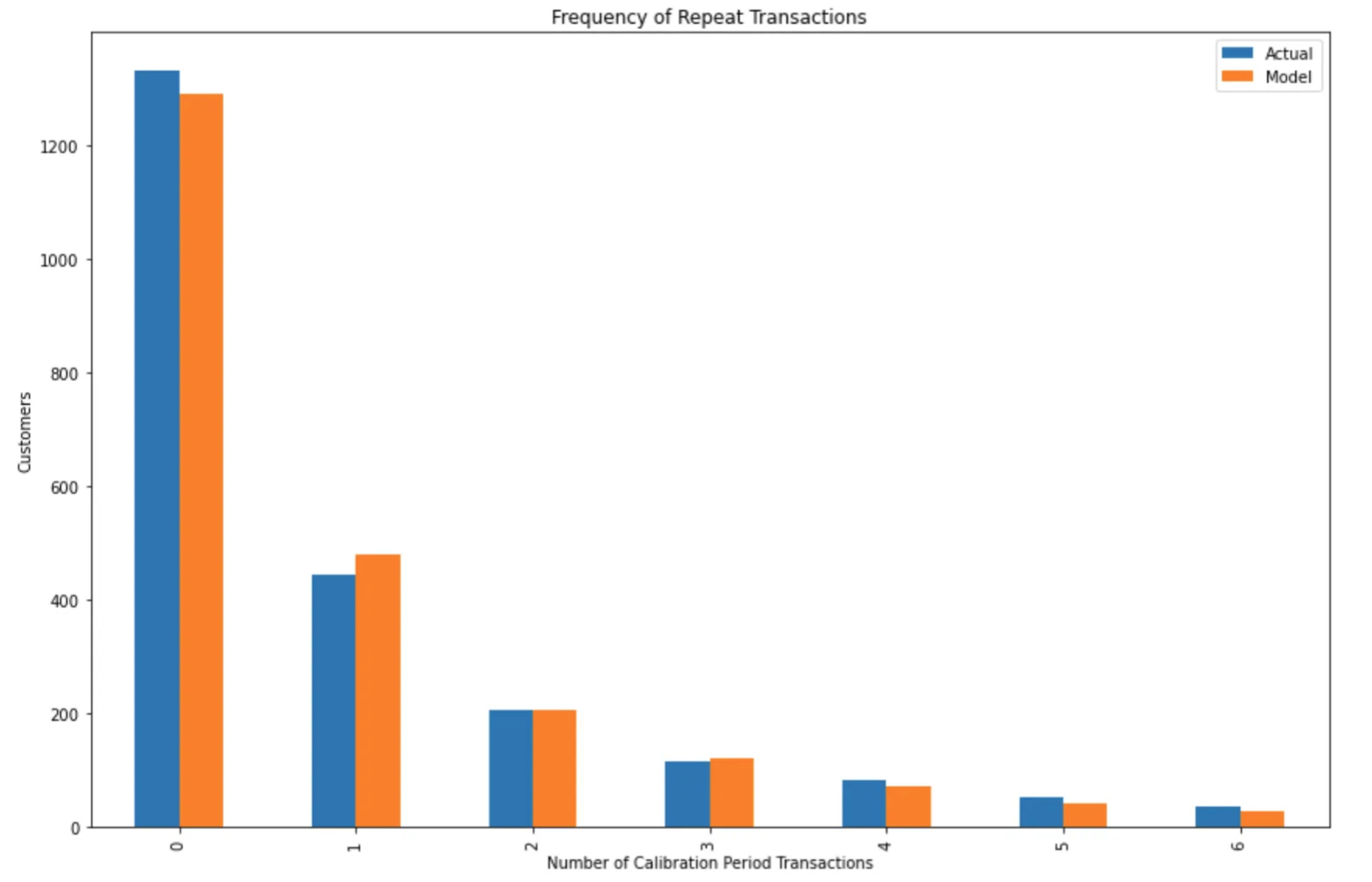}
      \label{fig:freq_repeat_trans_2}
      \caption{Frequency of Repeat Transactions (Model's prediction vs Actual)}
\end{figure}

The second is to split the dataset into two parts - Calibration and Holdout Datasets. For our case, we kept Jan 2022 - May 2022 as Calibration Dataset and May 2022 - July 2022 as Holdout Dataset. The calibration-holdout split divides our transactions into two depending on whether they fall into the calibration period or the observation period. The transactions that took place during the calibration period are used to train the model, whereas those occurring during the observation period (“holdout’” transactions) are used to validate the model. (Figure 2)

\begin{figure}[thpb]
      \centering
      \includegraphics[width=\linewidth]{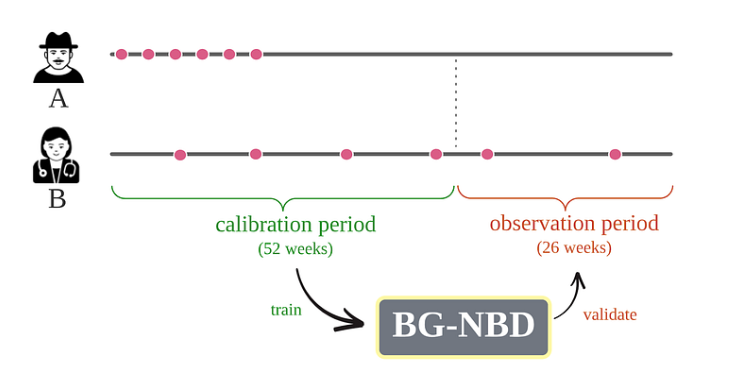}
      \label{fig:calibration_holdout_avatar}
      \caption{Use of Calibration and Holdout Dataset for validating the BGNBD model}
\end{figure}

With this split, we can actually see the distribution of the Actual Purchases vs the Model's prediction to get an estimate of the model's goodness for the holdout and calibration period. (Figure 3)

\begin{figure}[thpb]
      \centering
      \includegraphics[width=\linewidth]{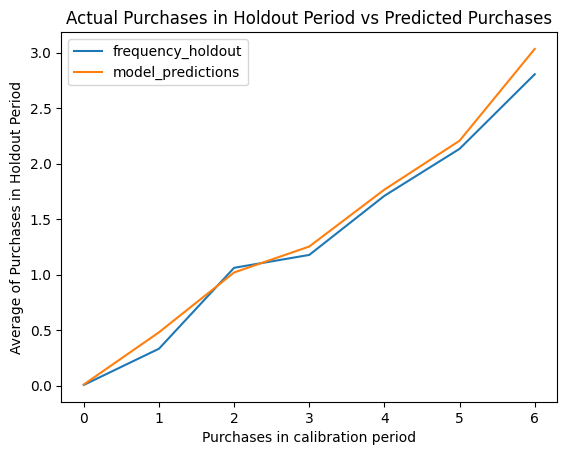}
      \label{fig:calibration_holdout_result}
      \caption{Actual Purchases vs Predicted Purchases}
\end{figure}

For the Holdout period, the different regression metrics are as follows:

\begin{center}
\begin{tabular}{||c c||} 
 \hline
 metric & value \\ [0.5ex] 
 \hline\hline
 Mean Squared Error & 1.25 \\ 
 \hline
 Mean Absolute Error & 1.1 \\
 \hline
 Mean Squared Log Error & 0.5 \\
 \hline
\end{tabular}
\end{center}

\subsection{Comparison of BGNBD and Gamma-Gamma Distribution with Other CLV Estimation Methods}

The BGNBD and Gamma-Gamma distribution models offer a robust and flexible approach to estimating Customer Lifetime Value (CLV). However, it is essential to compare their performance with other CLV estimation methods to better understand their advantages and limitations. This section discusses the comparison of the BGNBD and Gamma-Gamma models with other popular CLV estimation methods, such as traditional RFM (Recency, Frequency, Monetary) models and the Pareto/NBD model.

BGNBD and Gamma-Gamma vs. Traditional RFM Models: Traditional RFM models segment customers based on their recency, frequency, and monetary value scores. While these models provide a simple and intuitive way to analyze customer behavior, they lack the probabilistic and predictive capabilities of the BGNBD and Gamma-Gamma models. The BGNBD and Gamma-Gamma models account for the heterogeneity in customer behavior and can estimate the expected number of future transactions and monetary values, offering a more accurate and granular estimation of CLV.

BGNBD and Gamma-Gamma vs. Pareto/NBD Model: The Pareto/NBD model is another probabilistic model used for estimating CLV. While it shares similarities with the BGNBD model, the Pareto/NBD model assumes that the transaction process follows a Poisson process with a constant rate over time, whereas the BGNBD model allows for individual-specific transaction rates. Additionally, the BGNBD model uses a beta-geometric distribution for dropout (churn) probabilities, whereas the Pareto/NBD model uses a Pareto distribution. The BGNBD model generally provides a better fit to the data and more accurate predictions, especially when there is significant variation in customer transaction rates.

When comparing the BGNBD and Gamma-Gamma models with alternative CLV estimation methods, it is essential to consider factors such as the accuracy of predictions, the goodness of fit to the data, and the complexity of the models. The BGNBD and Gamma-Gamma models offer several advantages over traditional RFM and Pareto/NBD models in terms of their ability to account for the heterogeneity in customer behavior and provide more accurate predictions of future transactions and monetary values. However, their increased complexity may require more computational resources and a deeper understanding of the underlying statistical assumptions. As such, it is important to carefully consider the specific use case and data characteristics when selecting an appropriate CLV estimation method.

\subsection{Key Insights from the Analysis}

After evaluating the performance of the BGNBD and Gamma-Gamma distribution models and comparing them with other CLV estimation methods, several key insights can be drawn from the analysis. These insights can help guide marketing and retention strategies, improve customer engagement, and ultimately increase the overall profitability of a business. Some of the key insights derived from the analysis are as follows:

\begin{enumerate}
    \item \textbf{Identification of High-Value Customers:} The BGNBD and Gamma-Gamma models allow for the identification of high-value customers, who exhibit a higher expected number of transactions and monetary value. By targeting these customers with personalized offers, promotions, and retention strategies, businesses can improve customer loyalty, increase sales, and maximize the return on investment for their marketing efforts. (Figure 4)
    
    \begin{figure}[thpb]
          \centering
          \includegraphics[width=\linewidth]{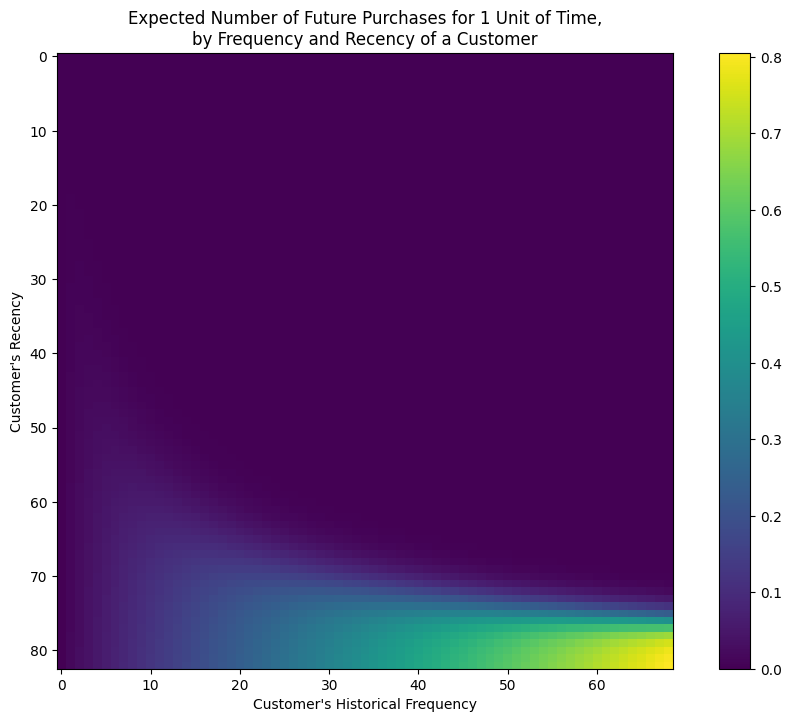}
          \label{fig:expected_purchase_heatmap}
          \caption{Expected number of purchases in 1 unit of time by Frequency and Recency} 
    \end{figure}
    In Figure 4, the x-axis represents the frequency of purchases, while the y-axis represents the recency of the most recent purchase. The color gradient in the plot indicates the probability of a customer being alive, with bluish colors representing lower expected transactions and greener colors indicating higher expected transactions.    
    The Frequency-Recency Matrix plot can reveal several important patterns and trends:
    \begin{itemize}
        \item Customers with higher frequency and recency values tend to have a higher expected number of future transactions. This means that customers who have made more purchases and have purchased more recently are more likely to make future purchases, which aligns with general customer behavior expectations.
        \item As frequency increases but recency remains low, the expected number of future transactions decreases. This indicates that customers who have not made a purchase recently, even if they have a high purchase frequency, are less likely to make future purchases.
        \item Conversely, customers with low frequency but high recency values may still have a relatively high expected number of future transactions. This suggests that recent purchases are an important factor in determining the likelihood of a customer making future purchases, even if they have not made many purchases overall.
    \end{itemize}

    The Frequency-Recency Matrix plot can be a useful tool for visualizing the relationship between customer recency, frequency, and expected future transactions. This information can help businesses develop targeted marketing and retention strategies to maintain relationships with active customers and encourage more frequent purchases from less active customers.
    
    \item \textbf{Heterogeneity in Customer Behavior:} The BGNBD and Gamma-Gamma models account for the inherent heterogeneity in customer transaction frequency, recency, and monetary value. By understanding these differences, businesses can tailor their marketing and retention strategies to cater to the unique preferences and behaviors of different customer segments, ultimately leading to more effective and efficient use of marketing resources. (Figure 5)
    \begin{figure}[thpb]
          \centering
          \includegraphics[width=\linewidth]{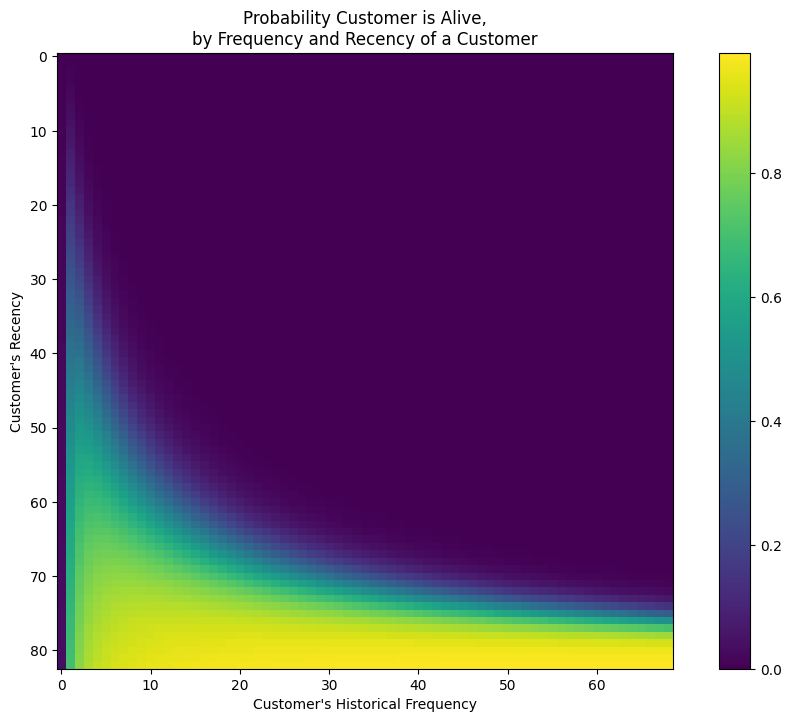}
          \label{fig:probability_alive_heatmap}
          \caption{Probability that a User is alive by Frequency and Recency} 
    \end{figure}

    In Figure 5, the x-axis represents the frequency of purchases, while the y-axis represents the recency of the most recent purchase. The color gradient in the plot indicates the probability of a customer being alive, with bluish colors representing lower probabilities and greensish colors indicating higher probabilities.

    The Probability Alive Matrix plot can reveal several important patterns and trends:

    \begin{itemize}
        \item Customers with higher frequency and recency values tend to have a higher probability of being alive. This means that customers who have made more purchases and have purchased more recently are more likely to be active customers, which aligns with general customer behavior expectations.
        \item As frequency increases but recency remains low, the probability of being alive decreases. This indicates that customers who have not made a purchase recently, even if they have a high purchase frequency, are less likely to be active customers.
        \item Conversely, customers with low frequency but high recency values may still have a relatively high probability of being alive. This suggests that recent purchases are an important factor in determining the likelihood of a customer being active, even if they have not made many purchases overall.
    \end{itemize}

    The Probability Alive Matrix plot can be a useful tool for visualizing the relationship between customer recency, frequency, and their likelihood of being active customers. This information can help businesses develop targeted marketing and retention strategies to re-engage inactive customers and maintain relationships with active customers.

    \item \textbf{Churn Prediction and Retention Strategies:} The BGNBD model provides an estimation of the probability of a customer becoming inactive (churning) after each transaction. This information can be used to identify customers at risk of churning and develop targeted retention strategies aimed at re-engaging these customers and encouraging repeat purchases. (Figure 6 and 7)

    \begin{figure}[thpb]
          \centering
          \includegraphics[width=\linewidth]{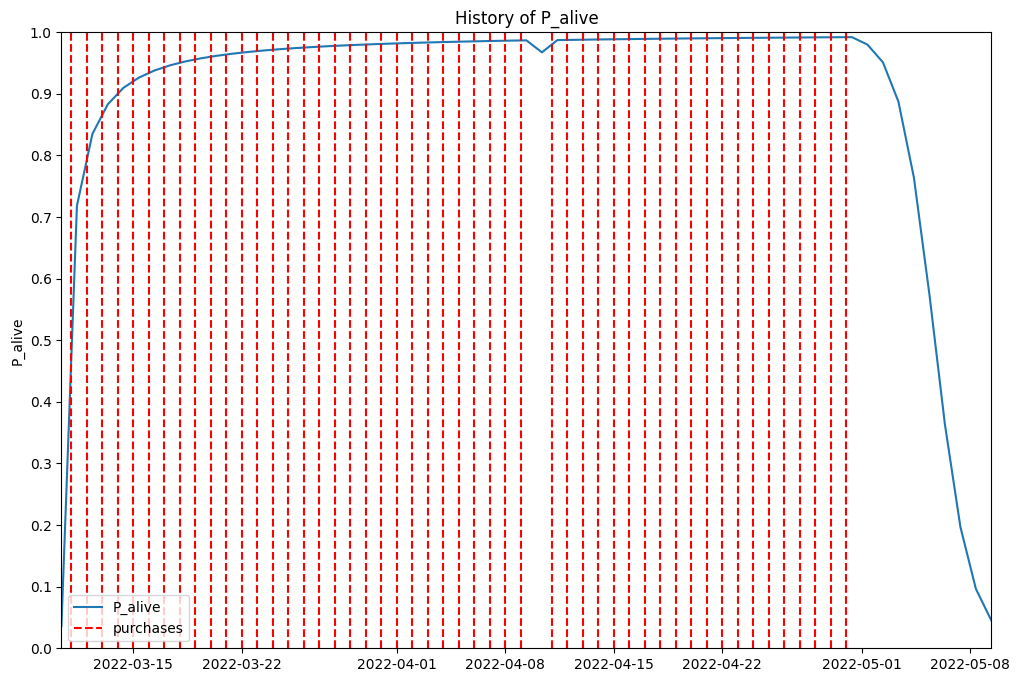}
          \label{fig:user_journey_1}
          \caption{User 1 Churn Timeline} 
    \end{figure}

    \begin{figure}[thpb]
          \centering
          \includegraphics[width=\linewidth]{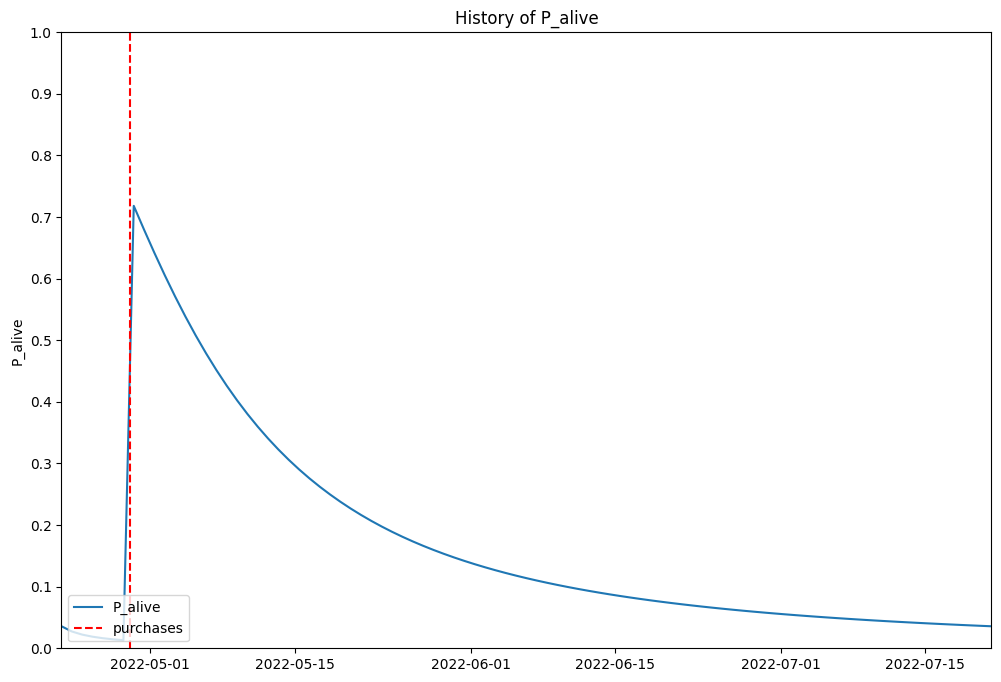}
          \label{fig:user_journey_1}
          \caption{User 2 Churn Timeline} 
    \end{figure}

    Figure 6 depicts a user with a high frequency. This user has a recency of 60 days (i.e. this user's first transaction was 60 days ago). Figure 7 depicts a user with a low frequency, but a higher recency of 90 days.

    It is evident from Figure 6, that even though User 1 made a lot of transactions in a short amount of time, as this user has lower recency, the probability of this user being active drops sharply after the last transaction. This is in logical sync with Figure 5.

    On the other hand, User 2 in Figure 7, did not make a lot of transactions (i.e., lower frequency), but this user has been a more loyal user in terms of recency. Hence this user's probability of being active drops at a much lower rate than User 1.

    Based on each user's Churn Timeline plot, businesses can take proactive decisions to make them active with marketing tactics before they are lost forever.

    \item \textbf{Improved Forecasting and Budgeting:} By providing more accurate predictions of future customer transactions and monetary values, the BGNBD and Gamma-Gamma models can help businesses improve their forecasting and budgeting processes. This can lead to better allocation of resources, more efficient marketing campaigns, and ultimately higher profitability.

    \begin{figure}[thpb]
          \centering
          \includegraphics[width=\linewidth]{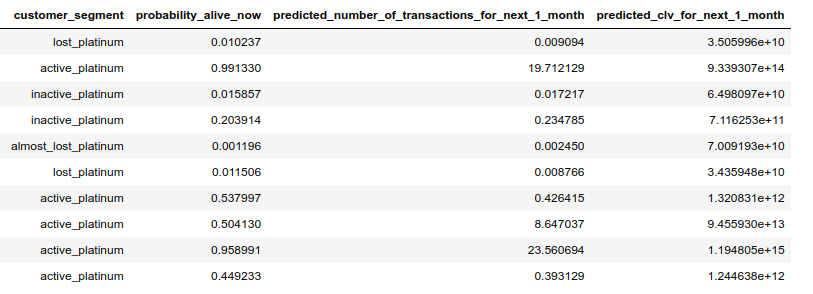}
          \label{fig:user_infos}
          \caption{Future Prediction of 10 random users} 
    \end{figure}

    Figure 8 depicts the model's prediction for 10 random users (User IDs concealed). So for each user, the model is able to predict:
    
    \begin{itemize}
        \item The user's probability of being active at that point in time.
        \item The expected number of transactions the user is going to make in the next 1 month.
        \item The expected amount of amount the user is going to spend (CLV) in the next 1 month.
    \end{itemize}
    
\end{enumerate}

In summary, the analysis of the BGNBD and Gamma-Gamma distribution models provides valuable insights into customer behavior, CLV estimation, and the effectiveness of different CLV models. These insights can be leveraged by businesses to inform their marketing and retention strategies, improve customer engagement, and ultimately drive higher profitability.

\section{Discussion}
\subsection{Implications for Marketing and Customer Retention Strategies}
The findings from the BGNBD and Gamma-Gamma distribution models' analysis provide valuable guidance for shaping marketing and customer retention strategies. By incorporating these insights, businesses can optimize their approach to customer engagement and retention. Some key implications include:

\begin{enumerate}
    \item \textbf{Tiered Rewards and Loyalty Programs:} With the ability to segment customers based on their lifetime value, businesses can develop tiered rewards and loyalty programs. By offering exclusive benefits and personalized incentives to high-value customers, businesses can strengthen their relationships with these customers, ultimately increasing loyalty and customer lifetime value.
    \item \textbf{Win-Back Campaigns:} Identifying customers who are at risk of churning allows businesses to design win-back campaigns targeting these specific individuals. By offering personalized promotions, discounts, or reminders of the value they derive from the product or service, businesses can potentially re-engage these customers and prevent them from churning.
    \item \textbf{Tailored Communication Strategies:} Understanding customer heterogeneity can inform tailored communication strategies. By segmenting customers based on their frequency and recency of purchases, businesses can create targeted messaging that resonates with specific customer groups. This approach ensures that customers receive relevant and timely information, improving engagement and conversion rates.
    \item \textbf{Cross-Selling and Up-Selling Opportunities:} By analyzing the monetary value of customers, businesses can identify potential cross-selling and up-selling opportunities. For high-value customers, businesses can suggest complementary products or services or offer premium versions of their existing products. These efforts can increase overall revenue and enhance customer satisfaction.
    \item \textbf{Data-Driven Decision Making:} The insights generated from the BGNBD and Gamma-Gamma distribution models can guide data-driven decision-making for marketing and customer retention initiatives. By using these models to make informed decisions, businesses can continuously refine their strategies and improve their ability to predict and react to customer behavior.
\end{enumerate}

In summary, the analysis of the BGNBD and Gamma-Gamma distribution models can inform various aspects of marketing and customer retention strategies. By leveraging these insights to create tiered rewards programs, win-back campaigns, tailored communication strategies, and targeted cross-selling and up-selling opportunities, businesses can effectively engage and retain their customer base, ultimately driving higher profitability.

\subsection{Limitations of the BGNBD and Gamma-Gamma Distribution Models}

While the BGNBD and Gamma-Gamma distribution models offer valuable insights into customer behavior and lifetime value estimation, they are not without limitations. It is important for businesses to be aware of these limitations when implementing these models in their marketing and customer retention strategies. Some key limitations include:

\begin{enumerate}
    \item \textbf{Assumptions about Customer Behavior:} The BGNBD model assumes that customer behavior follows a Poisson process, and the Gamma-Gamma model assumes that transaction values are independent of transaction frequency. However, in practice, customer behavior may not adhere to these assumptions, leading to inaccurate predictions or estimations.
    \item \textbf{Data Requirements:} Both the BGNBD and Gamma-Gamma models require sufficient historical data on customers' purchase frequency, recency, and monetary value to provide accurate predictions. For businesses with limited historical data or rapidly changing customer behavior patterns, these models may not produce reliable estimates.
    \item \textbf{Sensitivity to Model Parameters:} The performance of the BGNBD and Gamma-Gamma models is highly sensitive to the values of their parameters. If these parameters are not estimated accurately, the models can produce misleading predictions or estimations. Regular calibration and validation of model parameters are essential for maintaining their accuracy and reliability.
    \item \textbf{Lack of External Factors Consideration:} The BGNBD and Gamma-Gamma models do not explicitly account for external factors, such as economic conditions, competition, or changes in consumer preferences, which can significantly impact customer behavior and lifetime value. Incorporating these factors into the models may improve their predictive accuracy.
    \item \textbf{Model Complexity:} Although the BGNBD and Gamma-Gamma models offer a more sophisticated approach to CLV estimation compared to simpler methods, their complexity can be a barrier to implementation and interpretation for some businesses. Additionally, the computational cost of fitting these models may be prohibitive for large datasets or real-time applications.
\end{enumerate}

In summary, while the BGNBD and Gamma-Gamma distribution models provide valuable insights and improved accuracy in estimating customer lifetime value, it is essential for businesses to consider their limitations. To overcome these limitations, businesses should regularly calibrate and validate model parameters, consider external factors, and explore complementary approaches to improve the overall accuracy and reliability of their CLV estimates.

\subsection{Future Research Directions}
To further enhance the understanding of customer behavior and improve the accuracy of customer lifetime value estimation, several avenues for future research can be explored. Some potential future research directions include:

\begin{enumerate}
    \item \textbf{Integration of External Factors:} Investigating how external factors, such as economic conditions, competition, and changes in consumer preferences, impact customer behavior and lifetime value estimates can provide valuable insights. Future research could focus on incorporating these factors into the BGNBD and Gamma-Gamma models or developing new models that account for these influences.
    \item \textbf{Model Adaptation for Different Industries:} Investigating the performance of the BGNBD and Gamma-Gamma models across different industries can help uncover industry-specific insights and potential model modifications. This research can lead to the development of tailored models that are better suited for specific industries, improving the accuracy of CLV estimation.
    \item \textbf{Real-Time CLV Estimation:} Developing methods for real-time customer lifetime value estimation can enable businesses to make more informed, data-driven decisions in rapidly changing environments. Future research could focus on adapting the BGNBD and Gamma-Gamma models for real-time applications or exploring alternative approaches for real-time CLV estimation.
    \item \textbf{Incorporating Customer Engagement Metrics:} Investigating the relationship between customer engagement metrics, such as social media interactions or website visits, and customer lifetime value can provide valuable insights into the drivers of customer behavior. Future research could explore ways to incorporate these engagement metrics into the BGNBD and Gamma-Gamma models or develop new models that explicitly account for customer engagement.
\end{enumerate}

\section{Conclusion}
\subsection{Summary of Key Findings}

In conclusion, the application of the BGNBD and Gamma-Gamma distribution models in estimating customer lifetime value has demonstrated promising results. The key findings from the analysis and their implications for marketing and customer retention strategies are as follows:

\begin{enumerate}
    \item The BGNBD model effectively captures the heterogeneity in customer behavior, providing valuable insights into customers' purchase frequency, recency, and the probability of future transactions. The Gamma-Gamma model allows for the estimation of monetary value, further enhancing the understanding of customer lifetime value.
    \item The analysis revealed that customers with higher frequency and recency values tend to have higher expected future transactions, emphasizing the importance of maintaining customer engagement and loyalty.
    \item The BGNBD and Gamma-Gamma models outperformed traditional CLV estimation methods, offering improved accuracy and the ability to account for customer behavior heterogeneity.
    \item The insights derived from the analysis can be applied to inform marketing and customer retention strategies, such as tiered rewards programs, win-back campaigns, tailored communication strategies, and targeted cross-selling and up-selling opportunities.
    \item Despite their advantages, the BGNBD and Gamma-Gamma models have limitations, such as assumptions about customer behavior, data requirements, and sensitivity to model parameters. It is essential for businesses to consider these limitations when implementing these models and to explore future research directions to further enhance their accuracy and applicability.
\end{enumerate}

Overall, the BGNBD and Gamma-Gamma distribution models provide valuable tools for understanding customer behavior and estimating customer lifetime value. By leveraging these insights, businesses can optimize their marketing and customer retention strategies, ultimately driving increased customer engagement, loyalty, and profitability.

\subsection{Recommendations for Businesses and Practitioners}

Based on the findings and insights from the BGNBD and Gamma-Gamma distribution models, businesses and practitioners can implement a range of strategies to optimize their marketing and customer retention efforts. The following recommendations provide guidance on how to leverage these models effectively:

\begin{enumerate}
    \item \textbf{Invest in Data Collection and Analysis:} To utilize the BGNBD and Gamma-Gamma models effectively, businesses must invest in collecting and analyzing high-quality customer data. Regularly updating and maintaining customer data ensures the accuracy and reliability of the CLV estimates produced by these models.
    \item \textbf{Validate and Calibrate Models:} Periodically validating and calibrating the BGNBD and Gamma-Gamma models ensures that the model parameters remain accurate and that the predictions are reliable. Businesses should allocate resources to monitor model performance and update parameters as needed.
    \item \textbf{Segment Customers Based on CLV:} Businesses can use the CLV estimates from the BGNBD and Gamma-Gamma models to segment customers into different value tiers. By targeting high-value customers with personalized marketing and retention strategies, businesses can maximize their return on investment and foster long-term customer loyalty.
    \item \textbf{Monitor Customer Engagement Metrics:} Regularly monitoring customer engagement metrics, such as purchase frequency and recency, can help businesses identify changes in customer behavior and proactively address potential issues. By incorporating these metrics into the BGNBD and Gamma-Gamma models, businesses can improve their understanding of customer behavior and enhance their marketing and retention strategies.
    \item \textbf{Combine Models with Other CLV Estimation Methods:} While the BGNBD and Gamma-Gamma models offer valuable insights, businesses should consider combining these models with other CLV estimation methods, such as machine learning or deep learning techniques. By using a multi-method approach, businesses can improve the overall accuracy and reliability of their CLV estimates.
    \item 
\end{enumerate}

To summarize, businesses and practitioners can benefit significantly from incorporating the BGNBD and Gamma-Gamma distribution models into their marketing and customer retention strategies. By investing in data collection and analysis, validating and calibrating models, segmenting customers based on CLV, monitoring customer engagement metrics, combining models with other CLV estimation methods, and sharing insights across departments, businesses can optimize their marketing and customer retention efforts, ultimately driving increased customer engagement, loyalty, and profitability.

\bibliography{bib}
\bibliographystyle{ieeetr}

\end{document}